# Observation of Higher-Order Sideband Transitions and First-Order Sideband Rabi Oscillations in a Superconducting Flux Qubit Coupled to a SQUID Plasma Mode

Yoshihiro Shimazu, Masaki Takahashi, and Natsuki Okamura

*Department of Physics, Yokohama National University, Yokohama 240-8501, Japan*

We report results of spectroscopic measurements and time-domain measurements of a superconducting flux qubit. The dc superconducting quantum interference device (SQUID), used for readout of the qubit, and a shunt capacitor formed an LC resonator generating a SQUID plasma mode. Higher-order red and blue sidebands were observed in a simple measurement scheme because the resonant energy of the resonator, 600 MHz, was comparable to the thermal energy. We also observed Rabi oscillations on the carrier transition and the first-order sideband transitions. Because the qubit was coupled to a single arm of the dc SQUID, the qubit-SQUID coupling was significant at zero bias current, where these phenomena were observed. The ratios between the Rabi periods for the carrier transition and the sideband transitions are compared with those estimated from the coupling constant, which was separately determined. The result may be explained by assuming initial excitation of the resonator.

KEYWORDS: flux qubit, Josephson junction, sideband, circuit QED, quantum computation

## 1. Introduction

Work toward the development of future quantum information technology, including quantum simulators[1] and quantum computers,[2] has led to the extensive study of various types of quantum bits (qubits) over the last decade.[3] Compared with other types of qubits, superconducting qubits[4] have the advantages of scalability and strong interaction with external fields, which may allow fast and robust control and readout. Superconducting qubits easily couple to superconducting resonators, such as *LC* resonators,[5,6] coplanar waveguide resonators,[7] and 3D cavities.[8] The coupled systems, which are known as circuit quantum electrodynamics (QED) systems,[9] have been given much attention as important hybrid quantum circuits.[10] The resonators in these systems can be employed as a data bus to transfer quantum information between qubits. They are also used to read out the qubit state[8] and to build quantum gates,[11,12] such as controlled-NOT gates.

In this paper, we present experimental results from a three-Josephson-junction (3-JJ) flux qubit[13] coupled to a superconducting quantum interference device (SQUID) plasma mode[14,15] that is associated with a dc SQUID for readout of the qubit. This mode is equivalent to an *LC* resonance in which *L* is the Josephson inductance of the dc SQUID and *C* is the shunt capacitance of the SQUID. Since the intrinsic nonlinearity of the Josephson inductance is very small in typical experimental conditions, the SQUID plasma mode is well represented by a linear resonator. Quantum entanglement between a flux qubit and a resonator was first observed in the coupled system of a flux qubit and the SQUID plasma mode.[14]

For the sample we studied, the resonant energy of the resonator was comparable to the thermal energy. This enabled us to observe higher-order red- and blue-sideband transitions, in which the photon number changed by more than one, using a simple scheme. We note that such transitions in a superconducting qubit-resonator system have only been reported,[16] to the best of our knowledge, for a flux qubit coupled with a linear *LC* resonator. These transitions are similar to Raman transitions in atoms and molecules, wherein the low-energy modes involved in the transition can be vibrational and rotational modes. We also observed Rabi oscillations on the carrier transition and the first-order sideband transitions.[14] These observations were made at zero bias current. The significant qubit-resonator coupling in this bias condition is due to our sample geometry, in which the qubit was coupled to a single arm of the dc SQUID, in contrast to the conventionally studied flux qubits.[14,15]

We compared the ratios between the Rabi periods for the carrier transition and the sideband transitions with those estimated from the qubit-resonator coupling constant $g$, which was separately determined. The result can be explained by assuming a small number of excited photons in the initial state. We show that, under appropriate conditions, the analysis of sideband Rabi oscillations would allow the determination of the photon distribution in the resonator, which has been examined previously by spectroscopic means,[17] and the coupling constant $g$. In previous studies, $g$ has been estimated from vacuum Rabi splitting[7,18] or vacuum Rabi oscillations[5,19] in the resonant regime and from the dispersive shift[8] of a resonator frequency in the dispersive regime. The method for estimating $g$ on the basis of Rabi periods of the carrier transition and the sideband transitions should be very useful in the deep dispersive regime, where estimation based on the vacuum Rabi splitting, vacuum Rabi oscillations, or the dispersive shift is impossible.

## 2. Experimental Methods

Figures 1 (a) and (b) show a schematic and an optical micrograph of the sample. The 3-JJ flux qubit is galvanically connected to the readout dc SQUID. Electron-beam lithography was used for the fabrication. After fabricating the bottom plate of the shunt capacitors, made from aluminum, its surface was oxidized in ambient air. Then the top plates of the shunt capacitors, the qubit, and the SQUID were fabricated. The two shunt capacitors with an area of ~4000 $\mu m^2$ each were connected in series. The shunt capacitors and the dc SQUID were linked by superconducting wires with a total length of 400 μm and a width of 0.3 μm. The small Josephson junctions were formed using shadow deposition of two films of aluminum, with a thickness of 25 nm each. The areas of two of the junctions in the qubit were approximately 0.03 $\mu m^2$ each, and the third junction was smaller by a factor of 0.8. The areas of the junctions of the SQUID were also approximately 0.03 $\mu m^2$. The dimensions of the loop of the qubit were 25 μm by 2.5 μm.

The qubit is dominantly coupled to a single arm of the dc SQUID,[20] in contrast to the conventionally studied flux qubits,[14,15,21] wherein the qubit is coupled to both arms of the dc SQUID symmetrically. We will show that the difference in the geometry has a significant effect on the qubit-SQUID coupling.

The measurement was performed at 20 mK, the base temperature of a dilution refrigerator. For spectroscopy and measurement of Rabi oscillations,[21] a bias current pulse consisting of a short pulse of duration 20 ns and a trailing plateau of duration 2.5 μs was applied to the SQUID immediately after a microwave pulse. Then the switching probability $P_{sw}$ of the SQUID was recorded, typically after 5000 trials. The electrical leads to the sample were carefully filtered, and magnetic shielding was provided, using a superconducting lead shield and a permalloy shield. A small superconducting magnet (diameter: 20 mm) was used to apply a magnetic flux to the sample.

## 3. Results and Discussion

The circulating current of the qubit changes its direction when $f_Q = \Phi_Q/\Phi_0$ crosses $0.5 + N$, where $\Phi_Q$ is the magnetic flux in the qubit loop, $\Phi_0 = h/2e$, and $N$ is an integer. This current reversal is exhibited by the step-like variation in $P_{sw}$ shown in Fig. 2(a). In this measurement, under irradiation of a 10-GHz microwave pulse with a duration of 400 ns, the height of the bias current pulse was changed linearly as a function of $f_Q$. For low microwave power, we observed resonant peaks and dips when the energy separation between the ground state and the first excited state of the qubit $h\nu_Q$ was equal to $mh\nu$ where $\nu$ is the microwave frequency and $m$ is an integer. The appearance of resonances corresponding to $m > 1$ is due to the possible multi-photon processes and the harmonics produced by the nonlinearity of the microwave mixers used to produce the pulses. The presence of the harmonics was confirmed using a spectrum analyzer. Figure 2(a) shows the resonant peaks and dips for $m = 2$ and 3 at $\nu = 10$ GHz.

Figure 2(b) shows the energy dispersion curve of the qubit, which was obtained by locating the resonant peak and dip for $m = 1$ as a function of the applied magnetic flux. This result fits well to the theoretical energy splitting,[13]

$$h\nu_Q = \sqrt{\varepsilon^2 + \Delta^2} , \tag{1}$$

where $\varepsilon = 2I_p\Phi_0(f_Q - 0.5)$, $I_p$ is the circulating current in the qubit, and $\Delta$ is the minimum energy splitting (gap). From this fit, we obtained the parameter values $I_p = 217$ nA and $\Delta = 15.4$ GHz.

At $f_Q = 0.4954$, where $\Delta E = 16.7$ GHz, $P_{sw}$ vs. $\nu$ curves were obtained at various microwave powers, as shown in Fig. 3. In this measurement, the SQUID bias current $I_b$ was zero during the microwave pulses, which were 100-ns long. One remarkable observation that was made is that sideband peaks gradually appear with increasing microwave power, as can be seen in the figure. These sideband peaks are attributed to red and blue sidebands in the coupled system of the qubit and resonator. The transition frequency for the sidebands is given by $\nu_Q + M\Omega$ where $M$ is an integer other than 0 and where $\Omega$ is the resonator frequency. The transition for the sidebands is schematically shown in Fig. 4. $M$ represents the change in the quantum number of the resonator, which is also referred to as the resonator photon number. In Fig. 3, the sideband transitions for $|M| = 2$ are clearly visible. Those for $|M| = 3$ are barely observed at the highest microwave power. The apparent downward shift in resonator frequency with increasing power can be attributed to the nonlinearity of the resonator.[22] Because $\Omega \sim 0.6$ GHz corresponds to 30 mK, which is near the cryogenic temperature of 20 mK, a significant amount of photons are thermally excited, in contrast to the case of the previous study.[14] These thermally excited photons are seeding the red-sideband generation process and leading to the well-defined appearance of the first and second red sidebands as shown in Fig. 3.

$\Omega$ was also spectroscopically determined when the frequency of the microwave pulse was varied around 0.6 GHz. From the dependence of the resonator frequency on the magnetic flux and $I_b$,[14,22] shown in Fig. 5, we conclude that the origin of the resonator mode is the SQUID plasma mode associated with the combined system of the readout dc SQUID and the shunt capacitance $C$. The eigenfrequency of the SQUID plasma mode is given by

$$\nu_{pl} = \frac{1}{2\pi\sqrt{(L_{SQ} + L_s)C}}, \quad (2)$$

where $L_{SQ}$ is the Josephson inductance of the dc SQUID and $L_s$ is the stray inductance. For a symmetric dc SQUID, $L_{SQ}$ is given by

$$L_{SQ} = \frac{\Phi_0}{2\pi\sqrt{(2I_c \cos(\pi f_{SQ}))^2 - I_b^2}}, \quad (3)$$

where $f_{SQ} = \Phi_{SQ}/\Phi_0$, $\Phi_{SQ}$ is the magnetic flux threading the SQUID loop, and $I_c$ is the critical current of the junction in the SQUID. The observed resonant frequency is compared with the theoretical curves for $\nu_{pl}$ in Fig. 5. Good agreement between them is found. Using $I_c = 0.2$ µA and the fit shown in Fig. 5, we obtain the values $L_s \simeq 2300$ pH and $C \simeq 17$ pF. We note that the possible asymmetry (~10%) of the SQUID junctions does not change these estimates significantly because the asymmetry has a second-order effect on $L_{SQ}$.

The coupling between the flux qubit and the SQUID plasma mode is described by the Hamiltonian[5]

$$H = \frac{h}{2}(\varepsilon\sigma_z + \Delta\sigma_x) + \frac{h}{2}\Omega\left(a^\dagger a + \frac{1}{2}\right) + h\{g_1(a^\dagger + a) + g_2(a^\dagger + a)^2\}\sigma_z, \quad (4)$$

where $\sigma_z$ and $\sigma_x$ are Pauli matrices written in the persistent current states basis and $a^\dagger$ ($a$) is the photon creation (annihilation) operator; also, $g_1 = \frac{1}{2}\frac{d\varepsilon}{dI_b}\delta i_0$, and $g_2 = \frac{1}{4}\frac{d\varepsilon^2}{d^2 I_b}\delta i_0^2$, where $\delta i_0 = \sqrt{\frac{h\Omega}{2(L_{SQ} + L_s)}}$ represents the rms fluctuations of the current in the oscillator ground state. When the qubit is coupled to both arms of the SQUID symmetrically, $g_1$ becomes zero at $I_b = I_b^* \simeq 0$.[15,23] For the sample under investigation, wherein the qubit is coupled to a single arm of the dc SQUID, the curve of $\varepsilon$ vs. $I_b$ is shown in Fig. 6. The data agree well with the parabolic fit.[15,23] The decoupling point ($g_1 = 0$) is given by $I_b^* \simeq -0.1$ µA, which is ~80% of the SQUID switching current. On the basis of Fig. 6, we find $g_1 = 100 \pm 20$ MHz at $I_b = 0$ using the above expression. This large coupling, $g_1/\Omega \simeq 0.2$, allowed the clear observation of the sideband transitions at $I_b = 0$.

We also found that $\Delta$ exhibited a monotonic dependence on $I_b$. The reason for this is not understood at present. However, this dependence was very weak, and thus does not affect the conclusions of the present paper.

The origin of the second-order sideband transitions ($\nu = \nu_Q \pm 2\Omega$) may be due to the nonlinear term (the $g_2$ term) in Eq. (4). However, even for a flux qubit coupled to a linear LC resonator, qubit transitions involving the exchange of up to 10 photons have been observed.[18] Further study is needed to fully understand the origin of the higher-order sideband transitions.

We observed Rabi oscillations for the carrier transition ($\nu = \nu_Q$) and the first blue- and red-sideband transitions ($\nu = \nu_Q \pm \Omega$),[14] as shown in Fig. 7. We now discuss these sideband Rabi oscillations. These data were again taken at $I_b = 0$ and $f_Q = 0.502$. These oscillations were measured in a different experimental run from that in which the spectroscopic data shown in Fig. 3 was taken. There is a slight difference in both $I_p$ and $\Delta$ between these measurements. The oscillations are well fitted by exponentially decaying sinusoids with a small exponentially changing background. From the fit, the Rabi periods for the carrier transition and the blue- and red-sideband transitions are found to be $\tau_{\text{center}} = 2.6 \pm 0.1$ ns, $\tau_{\text{blue}} = 29 \pm 2$ ns, and $\tau_{\text{red}} = 20 \pm 3$ ns, respectively. The decay time constants are 13–20 ns.

The ratio between these periods can be compared with the corresponding theoretical predictions. We take into account the excitation of photons in the resonator because the photon energy is comparable to the thermal energy for the sample we studied. The state vectors of the combined system of the qubit and the resonator with $n$ photons are denoted as $|g\ n\rangle$ and $|e\ n\rangle$ for the qubit in the ground state and in the excited state, respectively. The blue-sideband transition between $|g\ n\rangle$ and $|e\ n+1\rangle$ takes place with a period $\tau_{\text{blue}}(n)$, while the red-sideband transition between $|g\ n+1\rangle$ and $|e\ n\rangle$ takes place with a period $\tau_{\text{red}}(n)$ ($n = 0, 1, 2, \cdots$). The ratios between the periods are theoretically given by

$$\frac{\tau_{\text{blue}}(n)}{\tau_{\text{center}}} = \frac{\Omega(\nu_Q + \Omega)}{2\sqrt{n+1}g_1\nu_Q \cos\theta}, \tag{5}$$

$$\frac{\tau_{\text{red}}(n)}{\tau_{\text{center}}} = \frac{\Omega(\nu_Q - \Omega)}{2\sqrt{n+1}g_1\nu_Q \cos\theta}, \tag{6}$$

and $\quad \dfrac{\tau_{\text{blue}}(n)}{\tau_{\text{red}}(n)} = \dfrac{\nu_Q + \Omega}{\nu_Q - \Omega},\tag{7}$

where $\cos\theta = \varepsilon/\sqrt{\varepsilon^2 + \Delta^2}$ (see the Appendix).[24]

The observed ratio $\tau_{\text{blue}}/\tau_{\text{red}} = 1.5 \pm 0.3$ is larger than the theoretical value of 1.06 given by Eq. (7). We attribute this to the frequency dependence of the microwave attenuation of the cable; higher frequency microwaves are more strongly attenuated leading to slower Rabi oscillations.

At the operating point for the data shown in Fig. 7, we obtain $\cos\theta = 0.13 \pm 0.04$; then using Eqs. (5) and (6) and the previously estimated value of $g_1 = 100 \pm 20$ MHz, we find

$$\frac{\tau_{\text{blue}}(n)}{\tau_{\text{center}}} \simeq \frac{\tau_{\text{red}}(n)}{\tau_{\text{center}}} = (21 \pm 8) \times \frac{1}{\sqrt{n+1}}.$$

Note that $\nu_Q \gg \Omega$ here. Despite the large uncertainty, we can conclude that the observed ratios

$$\frac{\tau_{\text{blue}}}{\tau_{\text{center}}} = 11 \pm 1, \quad \frac{\tau_{\text{red}}}{\tau_{\text{center}}} = 8 \pm 2$$

are not consistent with the theoretical values for $n = 0$, but are consistent with those for $n \simeq 3$. We note that the discrepancy between the theoretical values for $n = 0$ and the observed ones cannot be explained in terms of the monotonic frequency dependence of the microwave attenuation.

The above result, implying the initial presence of a small number of photons, is reasonable, considering that the photon energy corresponding to 30 mK is comparable to the thermal energy of the system. However, the photon distribution cannot be determined on the basis of a straightforward comparison between the theoretical values and the observed ones for $\tau_{\text{blue}}/\tau_{\text{center}}$ and $\tau_{\text{red}}/\tau_{\text{center}}$. We found the relaxation time of the qubit to be $T_1 \simeq 100$ ns at $I_b = 0$ and $I_b = I_b^*$. The photon lifetime was $T_{\text{photon}} \simeq 100$ ns, which was derived from the width of the resonant peak of the SQUID plasma mode. The Ramsey-interference decay time was $T_{2\text{Ramsey}} \simeq 25$ ns at $I_b = I_b^*$. These time scales are comparable to the decay time of the sideband Rabi oscillations shown in Fig. 7. In this case, these relaxation processes can combine with sideband excitation to cause the sideband transitions with different values of $n$ to become mixed with each other. Therefore, to determine the photon distribution, the qubit relaxation, the dephasing of the qubit, and the lifetime of the photons should be taken into account. Further theoretical analysis is needed to simulate the observed sideband Rabi oscillations.

If the photon lifetime and qubit decoherence time are considerably longer than the periods of the sideband Rabi oscillations, the photon distribution function can be determined from the analysis of the sideband Rabi oscillations; after a time $t$ corresponding to the duration of the resonant blue-sideband pulse ($\nu = \nu_Q + \Omega$), the expectation value of $\sigma_z$, written in the energy eigenstates basis, is given by

$$<\sigma_z> = -\sum_{n=0}^{\infty} p(n)\cos(2\pi\Omega_{\text{blue}}\sqrt{n+1}\,t), \qquad (8)$$

where $p(n)$ is the probability for $n$ photons to be excited in the initial state and $\Omega_{\text{blue}} = \tau_{\text{blue}}(0)^{-1}$ is the Rabi frequency of the blue-sideband transition between |g 0> and |e 1>.[25,26] The red-sideband Rabi oscillations are described similarly with $\Omega_{\text{blue}}$ replaced by $\Omega_{\text{red}} = \tau_{\text{red}}(0)^{-1}$ (the Rabi frequency of the red-sideband transition between |g 1> and |e 0>). It should be noted that the Rabi oscillations given by Eq. (8) have the same form as the Rabi oscillations for a Rydberg atom interacting with a resonant cavity with photons with a probability distribution given by $p(n)$.[27] If photons are in the coherent state, collapse and revival of the Rabi oscillations may be observed as shown in Ref. 27. Meanwhile, if the photons obey a thermal distribution (Bose-Einstein statistics), the fit to Eq. (8) allows estimation of the effective temperature $T_{\text{eff}}$ of the system. Calculation assuming a thermal distribution with $T_{\text{eff}}$ comparable to the photon energy has shown that the dominant frequency of blue- (red-) sideband Rabi oscillations is given by $\Omega_{\text{blue}}(\Omega_{\text{red}})$ irrespective of $T_{\text{eff}}$. Therefore, our result cannot not explained in terms of Bose-Einstein distribution of photons. Deviation from the Bose-Einstein statistics and the probable peak of the photon distribution at around $n \simeq 3$ may be caused by possible environmental modes of 1–2 GHz.

When the photon energy is considerably higher than the thermal energy,[14] the initial excitation of the photons is negligible, and thus the blue-sideband transition between |g 0> and |e 1> should be observed. In this case, we can estimate the coupling constant $g$ from the Rabi periods for the carrier and the blue-sideband transition using Eq. (5) with $n = 0$, under the assumption that the photon lifetime and qubit decoherence time are sufficiently long. The frequency characteristics of the microwave line must be taken into account in this analysis. This method for estimating $g$ is particularly useful in the deep dispersive regime. It compliments the estimate of $g$ on the basis of vacuum Rabi splitting and vacuum Rabi oscillations in the resonant regime and that based on the dispersive shift of a resonator frequency in the dispersive regime. We note that the dispersive shift $g^2/\nu_Q$ for the system under investigation is only 0.6 MHz, which is too small to be observed.

Our observation of the resonator states involving many excitations of photons may provide a basis for future experiments manipulating many composite levels of the coupled system, including demonstration of quantum gates using sideband transitions[11,12,28] and dynamical cooling such as sideband cooling, which was realized in an ion-trap experiment.[29] Dynamical cooling using a tunable qubit[30] and realizing nonclassical states of a resonator[31] are also interesting.

The important characteristics of the SQUID plasma mode in comparison with resonators with a fixed resonant frequency, such as an LC resonator or a transmission line resonator, are the tunability of $\Omega$ and the intrinsic nonlinearity. Using the tunability of $\Omega$, we may realize resonance between the resonator and a 3-JJ flux qubit at the symmetry point.[5] The degree of nonlinearity of the SQUID plasma mode can be easily tuned via a bias current and an applied magnetic flux. By enhancing the nonlinearity considerably, the SQUID plasma mode could be used as a qubit system that is similar to the phase qubit.

## 4. Conclusion

In conclusion, we observed higher-order red- and blue-sideband transitions in a coupled system of a flux qubit and a SQUID plasma mode. The resonant energy of the SQUID plasma mode was comparable to the thermal energy, which allowed this observation to be made in a simple scheme. Rabi oscillations for the carrier transition and the first red- and blue-sideband transitions were also observed. The significant qubit-resonator coupling at $I_b = 0$, where these observations were made, was due to the sample geometry; the qubit was coupled to a single arm of the dc SQUID. We compared the ratios between the Rabi periods of the carrier transition and the sideband transitions, $\tau_{\text{blue}}/\tau_{\text{center}}$ and $\tau_{\text{red}}/\tau_{\text{center}}$, with those estimated from the qubit-resonator coupling constant $g$, which was separately determined from the dependence of the qubit flux bias on the bias current. The result may be explained by assuming the initial excitation of $n \simeq 3$ resonator photons. If the photon lifetime and qubit decoherence time are considerably longer than the sideband Rabi periods, the photon distribution function can be determined from the analysis of sideband Rabi oscillations. In addition, if the resonator energy is considerably higher than the thermal energy, $g$ can be estimated by examining the ratio of the Rabi periods. This method for estimating $g$ is useful, particularly in the deep dispersive regime.

**Acknowledgment**

This work was supported by a Grant-in-Aid for Scientific Research from the Japan Society for the Promotion of Science.

**Appendix**

In this Appendix, we present the effective Hamiltonian for the flux qubit coupled to the resonator with a linear coupling constant $g$.[5,6,24)] The flux penetrating the qubit loop is assumed to be modulated sinusoidally wih a frequency $\nu$. The Hamiltonian written in the qubit energy eigenstates basis is given by $H = H_{QR} + H_D$,

where

$$H_{QR} = \frac{h}{2}\nu_Q \sigma_z + h\Omega(a^\dagger a + \frac{1}{2}) + hg(\sigma_z \cos\theta - \sigma_x \sin\theta)(a^\dagger + a) \tag{A·1}$$

and the driving term is

$$H_D = \frac{A\cos 2\pi\nu t}{2}(\sigma_z \cos\theta - \sigma_x \sin\theta) ; \tag{A·2}$$

here, $\cos\theta = \varepsilon/\nu_Q$, and $A$ is the amplitude of the energy-bias modulation. The nonlinear term shown in Eq. (4) is neglected here because it is not relevant to the carrier transition and first-order sideband transitions, which are focused on in this Appendix. The eigenstates of $H_{QR}$ are referred to as the dressed states. We assume $g\cos\theta \ll \Omega$ and $g\sin\theta \ll |\nu_Q \pm \Omega|$.

The driving term written in the dressed states basis is

$$H'_D = \frac{A\cos 2\pi\nu t}{2}\{\sigma_z \cos\theta - (\sigma_+ + \sigma_-)\sin\theta\} + \frac{gA\nu_Q \sin^2\theta \cos 2\pi\nu t}{\nu_Q^2 - \Omega^2}\sigma_z(a^\dagger + a)$$

$$-\frac{gA\nu_Q \sin\theta \cos\theta \cos 2\pi\nu t}{\Omega(\nu_Q + \Omega)}(\sigma_+ a^\dagger + \sigma_- a) + \frac{gA\nu_Q \sin\theta \cos\theta \cos 2\pi\nu t}{\Omega(\nu_Q - \Omega)}(\sigma_+ a + \sigma_- a^\dagger) \tag{A·3}$$

to the first order of $g$, where $\sigma_\pm = \frac{1}{2}(\sigma_x \pm i\sigma_y)$.

In the interaction picture using the unperturbed Hamiltonian $H_0 = \frac{h}{2}\nu_Q \sigma_z + h\Omega(a^\dagger a + \frac{1}{2})$, $H'_D$ is transformed to

$$H''_D = \frac{A\cos 2\pi\nu t}{2}\{\sigma_z \cos\theta - (\sigma_+ e^{2\pi i\nu_Q t} + \sigma_- e^{-2\pi i\nu_Q t})\sin\theta\}$$

$$+\frac{gA\nu_Q \sin^2\theta \cos 2\pi\nu t}{\nu_Q^2 - \Omega^2}\sigma_z(a^\dagger e^{2\pi i\Omega t} + ae^{-2\pi i\Omega t})$$

$$-\frac{gA\nu_Q \sin\theta \cos\theta \cos 2\pi\nu t}{\Omega(\nu_Q + \Omega)}(\sigma_+ a^\dagger e^{2\pi i(\nu_Q+\Omega)t} + \sigma_- ae^{-2\pi i(\nu_Q+\Omega)t})$$

$$+\frac{gA\nu_Q \sin\theta \cos\theta \cos 2\pi\nu t}{\Omega(\nu_Q - \Omega)}(\sigma_+ ae^{2\pi i(\nu_Q-\Omega)t} + \sigma_- a^\dagger e^{-2\pi i(\nu_Q-\Omega)t}). \tag{A·4}$$

The ratios between the Rabi periods for the carrier transition ($\nu = \nu_Q$) and the first blue- and red-sideband transitions ($\nu = \nu_Q \pm \Omega$) shown in Eqs. (5)–(7) are obtained from Eq. (A·4) using the rotating-wave approximation.

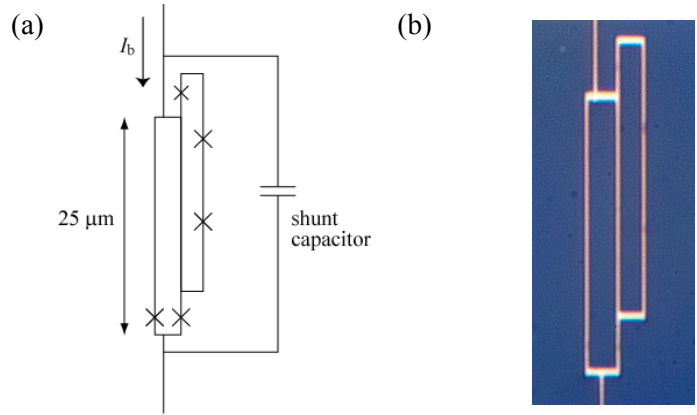

Fig. 1. (a) Schematic diagram of the sample. The crosses represent small Josephson junctions. The combination of the shunt capacitor and the dc SQUID generates a SQUID plasma mode, which couples to the three-Josephson-junction flux qubit. The resonator frequency can be tuned via the bias current $I_b$ and the magnetic flux in the SQUID loop. (b) Optical image of the qubit galvanically connected to the SQUID.

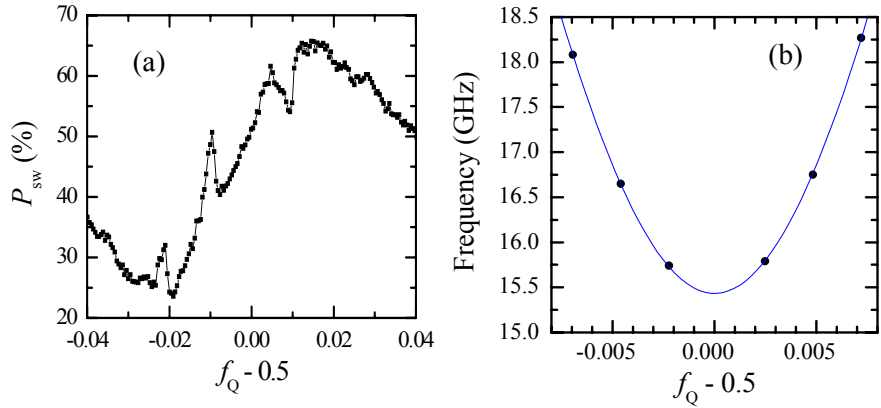

Fig. 2. (color online) (a) Switching probability $P_{sw}$ of the dc SQUID as a function of $f_Q - 0.5$ where $f_Q$ is the magnetic flux in the qubit loop divided by $\Phi_0 = h/2e$. A microwave pulse with frequency 10 GHz is applied to excite the qubit. Resonant peaks and dips are shown. (b) Qubit spectroscopy. The solid curve is a fit to the theoretical formula for the qubit frequency $\nu_Q$.

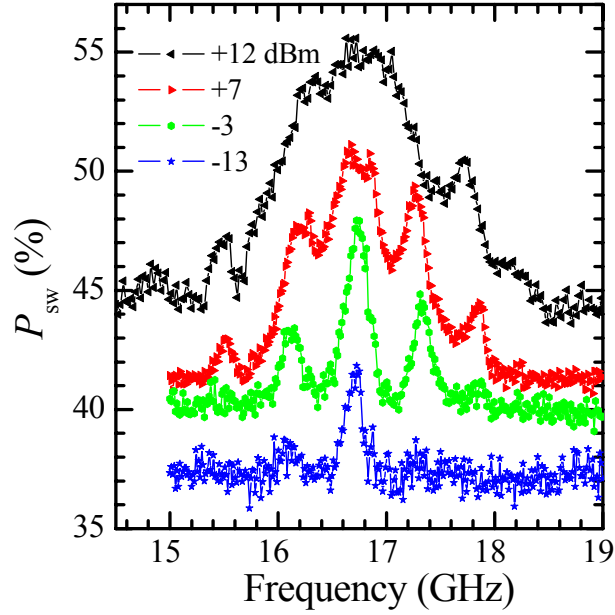

Fig. 3. (color online) Power dependence of $P_{sw}$ as a function of the microwave excitation frequency, measured at $f_Q = 0.4954$ and $I_b = 0$. The traces are offset vertically for clarity. With increasing power, blue- and red-sideband transitions of higher orders appear beside the carrier transition.

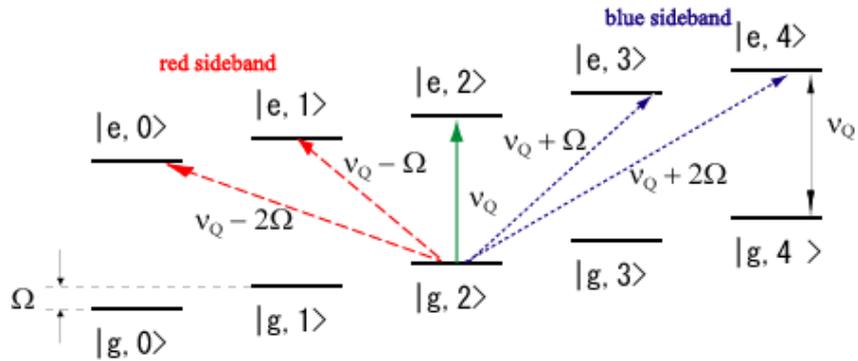

Fig. 4. (color online) Energy-level diagram for the qubit-resonator system. |g n> and |e n> denote the qubit in the ground state and in the excited state, respectively, accompanied by $n$ photons. The resonator energy, $h\Omega$, is considerably smaller than the excitation energy of the qubit, $h\nu_Q$. The carrier transition from |g 2> is shown by the green solid line, while the red- and blue-sideband transitions (of first and second order) are represented by red dashed and blue dotted lines, respectively.

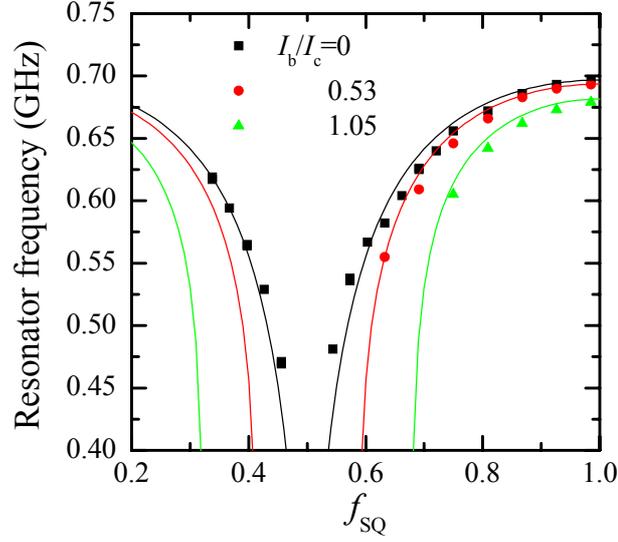

Fig. 5. (color online) The resonator frequency, that is, the center frequency of the resonant peak for the SQUID plasma mode, as a function of $f_{SQ} = \Phi_{SQ}/\Phi_0$, where $\Phi_{SQ}$ is the magnetic flux threading the SQUID loop, for different values of $I_b$. The solid lines are fits to the theoretical expression for the resonator frequency. See text for the parameters used for the fits.

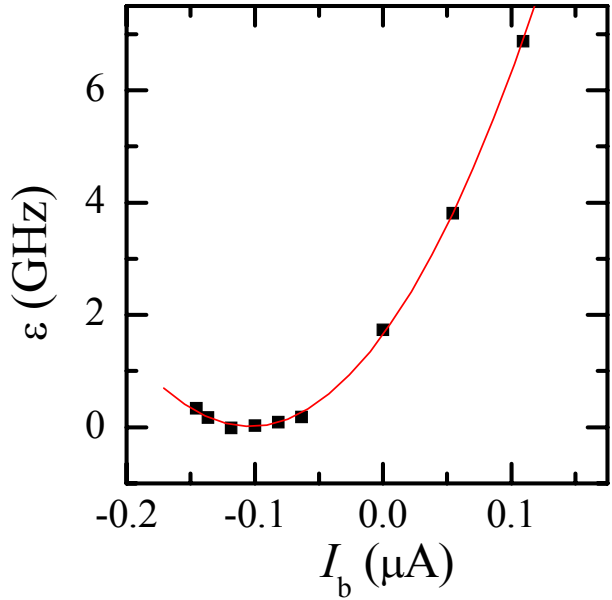

Fig. 6. (color online) Qubit energy bias $\varepsilon$ induced by $I_b$. The origin of $\varepsilon$ is chosen at its minimum. The linear coupling constant $g_1$, which is proportional to $d\varepsilon/dI_b$, becomes zero at $I_b = I_b^* \simeq -0.1$ μA. The solid line is a parabolic fit.

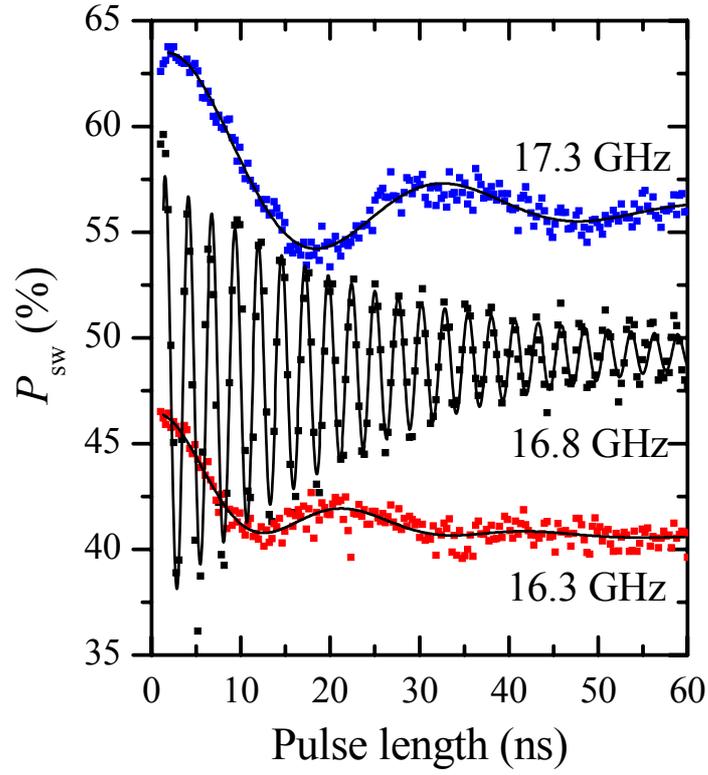

Fig. 7. (color online) Rabi oscillations for the carrier transition at 16.8 GHz and blue- and red-sideband transitions at 17.3 GHz and 16.3 GHz, respectively, measured at $f_Q = 0.502$ and $I_b = 0$. The black solid lines are fits to exponentially decaying sinusoids. The ratios between the Rabi periods can be discussed in terms of the qubit-resonator coupling constant.